\documentclass[pra,superscriptaddress,longbibliography,twocolumn,floatfix,showpacs,preprintnumbers,amsmath,amssymb,amsfonts,lengthcheck]{revtex4-1}

\usepackage{packages}  
\usepackage{commands}

\begin{document}

\title{Strong coupling of a chiral molecule with circularly polarised modes inside a cavity} 

\author{Lara Marie Tomasch}
\affiliation{\kassel}
\author{Fabian Spallek}
\affiliation{\kassel}
\author{Guido W. Fuchs}
\affiliation{\kassel}
\author{Thomas F. Giesen}
\affiliation{\kassel}
\author{Stefan Yoshi Buhmann}
\affiliation{\kassel}

\newcommand{\kassel}{Institut f\"{u}r Physik, Universit\"{a}t Kassel, Heinrich-Plett-Stra{\ss}e 40, 34132 Kassel, Germany}

\date{\today}

\begin{abstract}
We consider the discriminatory interaction of a chiral molecule with circularly polarised modes inside a cavity. Starting from a generalised Jaynes--Cummings model that includes both electric and magnetic dipole couplings, we derive the Rabi frequency and associated Casimir--Polder potential for a cavity with a single mode of given handedness. One finds that both acquire a discriminatory component whose sign depends on the relative handedness of molecule and cavity mode. We generalise this result to a cavity that supports two modes of different handedness on the basis of superradiant collective field states.
\end{abstract}

\maketitle


\section{Introduction}
\label{Sec:Introduction}

In its simplest form, cavity quantum electrodynamics (QED) is concerned with the interaction of a two-level atom with a single standing-wave mode of a cavity, as described by the paradigmatic Jaynes--Cummings model \cite{Jaynes63}. When the atom--field interaction dominates over the atomic and cavity loss channels, the model predicts Rabi oscillations in the form of reversible emission and reabsorption of cavity photons by the atom. From a spectroscopic point of view, this strong coupling is associated with the formation of pairs of atom--field states whose energy splitting is governed by the Rabi frequency. Rabi oscillations and many related quantum optics phenomena have been implemented in the pioneering experiments of Haroche  \cite{Haroche85} and many others, see, e.g.\ Refs.~\cite{Haroche89,Miller05} for reviews. Recently, cavity QED has been combined with density-functional theory in order to describe strong coupling of matter  systems of ever increasing complexity with quantized field modes \cite{Flick17,Baranov18,Riso22}.    

In the weak-coupling, perturbative limit, the interaction of a single particle with the vacuum electromagnetic field in a nontrivial environment gives rise to a position-dependent energy shift \cite{Casimir48}. It can be interpreted as a potential energy from which the Casimir--Polder force can be obtained. This idea can be generalised to the strong-coupling case within the framework of the Jaynes--Cummings model \cite{Jaynes63}, where the position-dependent vacuum Rabi frequency of a particle inside a cavity acts as a potential \cite{Haroche91, Englert91, Buhmann08}. Assuming that an initially excited particle enters an empty cavity adiabatically, the resulting force is opposite to the optical force induced by a classical laser: in the cavity-QED case, a positive atom--field detuning is associated with high-field seeking. 

The recent fabrication of chiral optical cavities that support light of a given circular polarisation \cite{Feis20,Kang20,Voronin22,Bassler24} has been exploited in the proposal of new schemes in an emerging chiral cavity QED \cite{Schafer23, Baranov23, Riso23}. Starting from the molecule-field coupling in multipolar form, an effective Tavis--Cummings model \cite{Tavis68, Tavis69} was developed for the interaction of an ensemble of chiral molecules with a cavity mode of single handedness \cite{Schafer23}. It was shown that the collective, superradiance-type strong coupling between the molecules and the mode may lead to discriminatory polaritonic ground-state energies that scale with the square root of the number of molecules. It was further revealed that chiral cavities can also induce discriminatory effects in the rotational spectrum of molecules \cite{Riso23} and it was suggested that chiral cavities may induce stereoselectivity \cite{Riso24}. As a foundation for chiral polaritonics, chiral cavity QED has been systematically developed for planar geometries \cite{Mauro24}. Compared to other enantio-selective techniques, e.g., photo-electron circular dichroism \cite{Tia17, Fehre21} or cold target recoil ion momentum spectroscopy \cite{Fehre22}, the use of a chiral optical cavity is a non-destructive technique that leaves the molecular structure and electron configuration unchanged.  

In this work, we are going to consider the strong coupling of a single chiral molecule with a chiral cavity to derive discriminatory Rabi frequencies and Casimir--Polder forces. In order to establish the basic setup and formalism, we start in Sect.~\ref{sec:SingleMode} with the simplest case of a cavity with a single circularly polarised mode. In Sect.~\ref{Sec:TwoMode}, we then study the more general case of a cavity with two modes of opposite handedness. A discussion of the resulting discriminatory Rabi frequencies and their typical order of magnitude in given in Sect.~\ref{Sec:Discussion}, followed by a conclusion in Sect.~\ref{Sec:Conclusion}.


\section{Single-mode cavity} 
\label{sec:SingleMode}

As a minimal chiral cavity scenario, we study the system Hamiltonian, Rabi frequency and Casimir--Polder force for a single-mode cavity.


\subsection{Model}

Our basic cavity-QED setup consists of a single chiral two-level molecule with ground state $\ket{g}$ and excited electronic state $\ket{e}$ and transition frequency $\omega_\mathrm{M}$ that interacts near-resonantly with a single circularly polarised mode of frequency $\omega$ and given handedness inside a chiral cavity. The Hamiltonian of this system 
consists of an molecular part, a field part and an interaction part,
\begin{equation}
    \hat{H}= \hat{H}_{\text{M}} + \hat{H}_{\text{F}} + \hat{H}_{\text{int}},
\end{equation} 
with
\begin{eqnarray}
    \hat{H}_{\text{M}}&=&\tfrac{1}{2}\hbar\omega_\mathrm{M}\hat{\sigma}_z,\\
    \hat{H}_{\text{F}}&=&\hbar\omega\hat{a}^\dagger\hat{a}
\end{eqnarray}
($\hat{\sigma}_z$: Pauli $z$-matrix, $\hat{a}^\dagger, \hat{a}$: mode creation and annihilation operators). In order to account for the optical activity of the molecule, the interaction part in multipolar coupling scheme and long-wavelength approximation \cite{Power59, Woolley71} contains both electric and magnetic dipole terms,
\begin{equation}
    \hat{H}_{\text{int}} = - \hat{\bm{d}}\cdot \hat{\bm{E}}- \hat{\bm{m}} \cdot \hat{\bm{B}}
\end{equation}

According to Lloyd's theorem \cite{Lloyd51}, the electric-dipole matrix element $\vec{d}=\bra{g}\hat{\vec{d}}\ket{e}$ and its magnetic-dipole counterpart $\vec{m}=\bra{g}\hat{\vec{m}}\ket{e}$ exhibit a relative phase $\pm\im$, so that we may write 
\begin{equation}
    \frac{\vec{m}}{c}=-\im\chi\vec{d}
\end{equation}
with a dimensionless real-valued chiral molecular parameter
\begin{equation}
\label{Eq:6}
    \chi=\frac{R}{c|\vec{d}|^2}\in[-1,1]
\end{equation}
whose sign and magnitude is governed by the molecule's (ground-state) rotatory strength $R=\mathrm{Im}(\vec{d}\cdot\vec{m}^\ast)$.

\begin{figure}
    \includegraphics[width=1.\columnwidth]{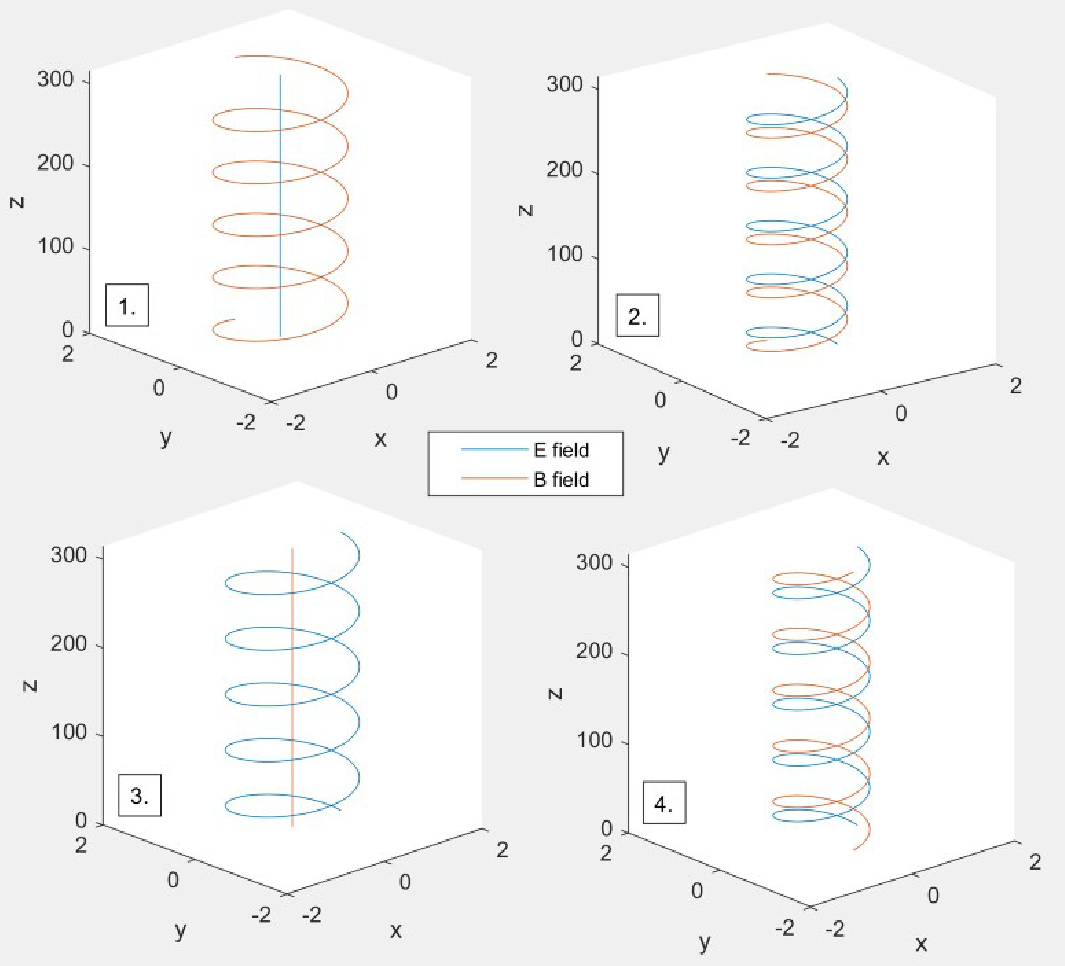}
    \caption{\textit{Circularly polarised cavity mode.} Illustration of  electric and magnetic fields at times $\omega t=0,\pi/2, \pi,3\pi/2$.}
    \label{Fig:1}
\end{figure}

In Coulomb gauge, the electric- and magnetic-field operators in the Heisenberg picture can be given in terms of the vector potential according to
\begin{equation}
\begin{split}
    \hat{\bm{E}} &= - \dot{\hat{\bm{A}}} \\
    \hat{\bm{B}} &= \bm{\nabla} \times \hat{\bm{A}}
\end{split}
\end{equation}
For a single circularly polarised mode, the latter reads
\begin{equation}
    \hat{\bm{A}}(\bm{r}, t)= \left[ \bm{A}(\bm{r}) \, \hat{a}\text{e}^{-\text{i}\omega t} + \bm{A}^*(\bm{r}) \, \hat{a}^\dagger \text{e}^{\text{i}\omega t} \right]
\end{equation}
with normalised mode function,
\begin{equation}
    \bm{A}(\bm{r})=A_0\vec{v}(\vec{r}),
\end{equation}
amplitude
\begin{equation}
\label{Eq:8}
    A_0=\sqrt{\frac{\hbar}{2\varepsilon_0\omega V}},
\end{equation}
and spatial profile
\begin{equation}
\label{Eq:9}
    \vec{v}(\vec{r})=
    \begin{pmatrix}
        \cos(kz)\\
        \mp\sin(kz)\\
        0
    \end{pmatrix}.
\end{equation}
Here, $V$ denotes the mode volume, the different signs refer to a right- or left-handed circularly polarised standing-wave mode, respectively, and we have assumed a planar cavity whose optical axis coincides with the $z$-axis. With these choices, the vector potential reads
\begin{equation}
    \hat{\bm{A}}(r, t) 
    = \left[ A_0  \bm{v}(\vec{r}) \, \hat{a} \, \text{e}^{-\text{i}\omega t} + A_0 \bm{v}(\vec{r}) \, \hat{a}^{\dagger} \, \text{e}^{\text{i}\omega t} \right],
\end{equation}
from which we can calculate the following field components,
\begin{equation}
\begin{split}
    \hat{\bm{E}}(\bm{r},t)&= \text{i} \omega A_0 \bm{v}(\vec{r}) \left[ \hat{a}\, \text{e}^{-\text{i}\omega t} - \hat{a}^\dagger  \, \text{e}^{\text{i}\omega t} \right], \\\
    \hat{\bm{B}}(\bm{r},t)&= \pm k A_0 \bm{v}(\vec{r}) \left[ \hat{a}\, \text{e}^{-\text{i}\omega t} + \hat{a}^\dagger  \, \text{e}^{\text{i}\omega t} \right].
\end{split}
\end{equation}
The position- and time-dependence of the electric and magnetic fields associated with a circularly standing-wave mode is illustrated in Fig.~\ref{Fig:1}. One sees that both fields are given by helical spatial profiles that oscillate in time ith a relative phase difference of $\pi/2$. In contrast to linearly polarised standing waves, the circularly polarised mode does not exhibit any spatial nodes where the fields vanish. 

Using the expressions for the electromagnetic field and returning to the Schr\"odinger picture, the interaction Hamiltonian assumes the form
\begin{equation}
    \hat{H}_\text{int}=  \hbar g \hat{\sigma}\hat{a}^{\dagger}  + \hbar g^*\hat{\sigma}^\dagger \hat{a} 
\end{equation}
with a coupling constant
\begin{multline}
    g=g(\vec{r})=\frac{\im\omega A_0}{\hbar}\vec{d}\cdot\vec{v}(\vec{r})\mp \frac{kA_0}{\hbar}\vec{m}\cdot\vec{v}(\vec{r})\\
    =\frac{\im\omega A_0}{\hbar}(1\pm\chi)\vec{d}\cdot\vec{v}(\vec{r})
\end{multline}
Here, we have applied the rotating-wave approximation by discarding the rapidly oscillating contributions $\hat{a}\hat{\sigma}$ and $\hat{a}^\dagger\hat{\sigma}^\dagger$. The interaction Hamiltonian is hence of Jaynes--Cummings form where the coupling constant exhibits a correction factor $(1\pm\chi)$ due to the magnetic-dipole interaction and its spatial dependence is governed by the circularly polarised mode~(\ref{Eq:8}) with (\ref{Eq:9}). In rotating-wave approximation, the total system Hamiltonian assumes a block diagonal form. In the $n+1$-excitation subspace spanned by the states $\ket{e}\ket{n},\ket{g}\ket{n+1}$, it can be given in matrix form
\begin{equation}
    H=\hbar\begin{pmatrix}
       n\omega+\tfrac{1}{2}\omega_\mathrm{M}&\sqrt{n+1}g^\ast\\[0.5ex]
       \sqrt{n+1}g&(n+1)\omega-\tfrac{1}{2}\omega_\mathrm{M}
    \end{pmatrix}.
\end{equation}
\vspace{0ex}


\subsection{Rabi frequency}

The eigenenergies of the system can be easily obtained by diagonalising the above matrix, yielding
\begin{equation}
\label{Eq:17}
    E_{1,2} = \bigl( n + \tfrac{1}{2} \bigr) \hbar \omega \pm \tfrac{1}{2} \hbar \, \Omega
\end{equation}
with Rabi frequency
\begin{equation}
\label{Eq:18}
    \Omega= \sqrt{\Delta^2 + 4(n+1){\lvert g \rvert}^2 }
\end{equation}
and molecule--field detuning $\Delta = \omega_\mathrm{M} - \omega$. The associated eigenstates are given by
\begin{eqnarray}
    \lvert E_1 \rangle &=& \cos(\theta)\lvert e \rangle \lvert n \rangle + \sin(\theta)\lvert g \rangle \lvert n+1 \rangle,\\
    \lvert E_2 \rangle &=& -\sin(\theta)\lvert e \rangle \lvert n \rangle + \cos(\theta)\lvert g \rangle \lvert n+1 \rangle,
\end{eqnarray}
with a coupling angle $\theta$ defined by
\begin{equation}
    \tan(2\theta)= - \frac{4(n+1)\vert g\vert^2}{\Delta} .
\end{equation}

The Rabi frequency depends on the handedness of both the molecule and the field mode via the coupling constant, where
\begin{equation}
    \lvert g \rvert^2= \frac{\omega^2A_0^2}{\hbar^2}|\vec{d}\sprod\vec{v}(\vec{r})|^2(1\pm\chi)^2.
\end{equation}
For a randomly oriented molecule, the rotational average $\overline{\vec{d}\otimes\vec{d}^\ast}=\tfrac{1}{3}|d|^2\tens{I}$ leads to the position-independent result
\begin{equation}
\label{Eq:20}
    {\vert g \vert }^2 =\frac{\omega^2A_0^2}{3\hbar^2} |d|^2(1\pm\chi)^2 .
\end{equation}
We see that the Rabi frequency is enhanced due to the magnetic coupling when the handedness of mode and molecule matches (i.e. $+$ for the mode and $\chi>0$ or $-$ for the mode and $\chi<0$). It is reduced when the molecule and the mode have opposite handedness (i.e. $+$ for the mode and $\chi<0$ or $-$ for the mode and $\chi>0$). 

\begin{figure}
    \includegraphics[width=1.\columnwidth]{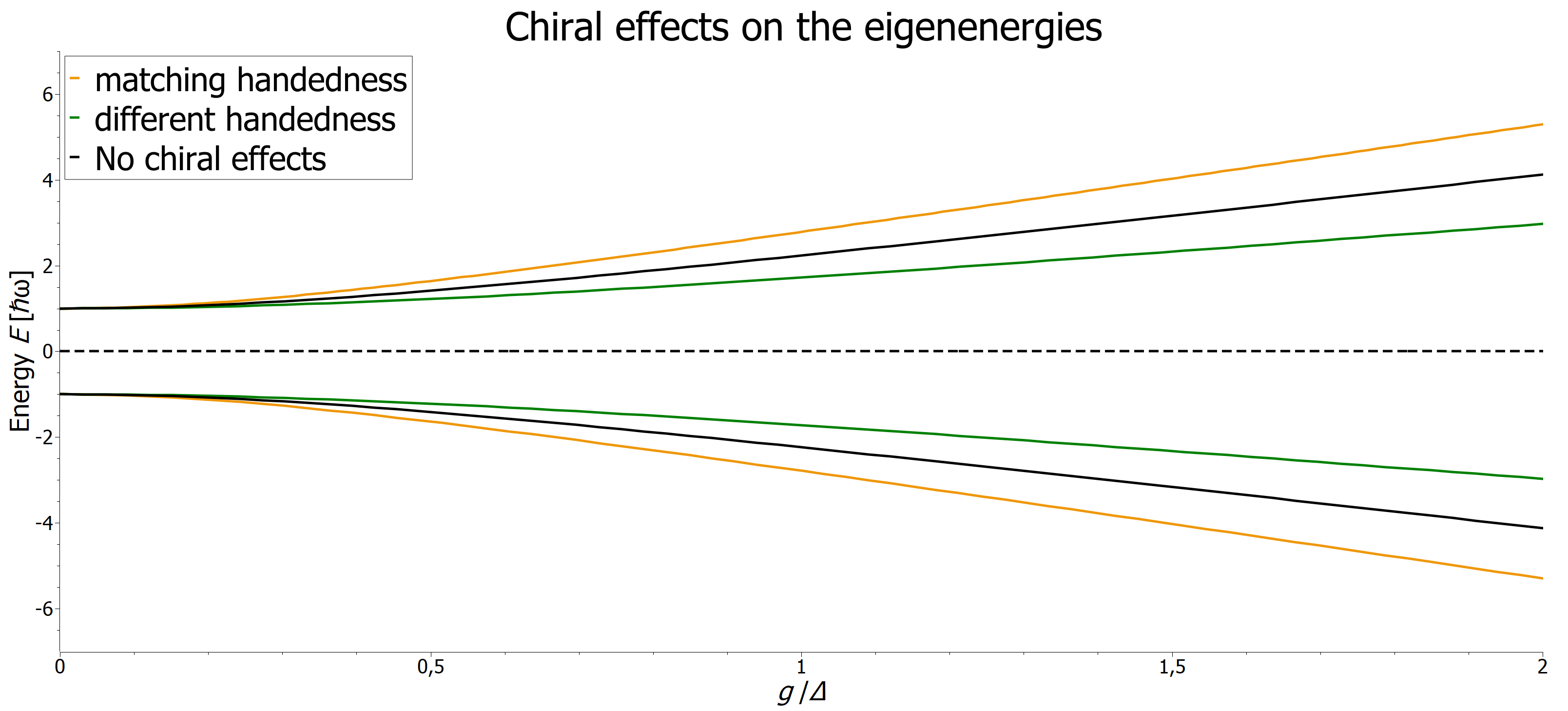}
    \caption{\textit{Chiral eigenenergies.} Illustration of the eigenenergies under influence of chirality.}
    \label{Fig:2}
\end{figure}


\subsection{Casimir--Polder force}

Following Refs.~\cite{Haroche91, Englert91, Buhmann08}, we are going to derive the strong-coupling Casimir--Polder force from the gradient of the eigenenergy. For a randomly oriented molecule interacting with a circularly polarised mode, the coupling constant~(\ref{Eq:20}) and hence also the Rabi frequency and eigenenergy are position-independent, so the Casimir--Polder force vanishes. Instead, we will consider a molecule with dipole moments oriented along the $x$-direction, where the coupling constant reads
\begin{equation}
     \lvert g \rvert^2(z)=  \frac{\omega^2A_0^2}{\hbar^2} |d_x|^2(1\pm\chi)^2\cos^2(kz).
\end{equation}
Assuming the system to be prepared in eigenstate $1$ and substituting the coupling constant into Eqs.~(\ref{Eq:17}) and (\ref{Eq:18}), we find a Casimir--Polder force directed along the $z$-axis, $\vec{F}=-\bm{\nabla}E_1=-(\mathrm{d}E_1/\mathrm{z})\vec{e}_z=F_z\vec{e}_z$ with
\begin{equation}
    F_z = \frac{2(n+1)k\omega^2A_0^2d_x^2(1\pm\chi)^2\cos(kz)\sin(kz)}{\sqrt{\Delta^2 + 4(n+1)\omega^2A_0^2d_x^2(1\pm\chi)^2\cos^2(kz)}}
\end{equation}
For zero detuning, $\Delta\to 0$, this expression simplifies to
\begin{equation}
    F_z=\hbar\sqrt{n+1} k\omega A_0|d_x| |1\pm\chi| \sin(kz)
\end{equation}
As in the case of the Rabi frequency, we find that the Casimir--Polder force is larger if the handedness of molecule and mode match due to the presence of the factor $|1\pm\chi|$.


\section{Two-mode cavity}
\label{Sec:TwoMode}

Next, we consider the case of a two-level molecule interacting with two circularly polarised standing-wave modes $1$ and $2$ inside a cavity, one left-handed and one right-handed. In straightforward generalisation of the single-mode case, the system Hamiltomian in rotating-wave approximation reads
\begin{multline}
    \hat{H} = \tfrac{1}{2}\hbar\omega_\mathrm{M} \hat{\sigma}_z + \hbar \omega_1 \hat{a_1}^\dagger \hat{a_1} + \hbar \omega_2\hat{a_2}^\dagger \hat{a_2}\\ + \hbar \bigl(g_1\hat{\sigma}\hat{a_1}^\dagger+ g_1^* \hat{\sigma}^\dagger\hat{a_1} \bigr) + \hbar \bigl( g_2 \hat{\sigma}\hat{a_2}^\dagger  + g_2^*\hat{\sigma}^\dagger \hat{a_2} \bigr)
\end{multline}
In the $n_1+n_2+1$-excitation sector as spanned by the states $\ket{e}\ket{n_1}\ket{n_2},\ket{g}\ket{n_1+1}\ket{n_2},\ket{g}\ket{n_1}\ket{n_2+1}$, the Hamiltonian assumes the matrix form
\begin{widetext}
\begin{equation}
\label{Eq:25}
  H = \hbar\begin{pmatrix} n_1\omega_1  + n_2\omega_2  + \tfrac{1}{2} \omega_\mathrm{M} & \sqrt{n_1 +1}g_1^*  &  \sqrt{n_2 +1}g_2^* \\ \sqrt{n_1 +1}g_1  & (n_1 +1)\omega_1  + n_2\omega_2  - \tfrac{1}{2} \omega_\mathrm{M} & 0 \\  \sqrt{n_2 +1}g_2 & 0 &  n_1\omega_1  + (n_2 +1)\omega_2  - \tfrac{1}{2} \omega_\mathrm{M} \\
\end{pmatrix}.
\end{equation}
\end{widetext}


\subsection{Degenerate case}

We first consider the degenerate case 
$\omega_1=\omega_2=\omega$, $g_1=g_2=g$, $n_1=n_2=n$ which applies for an achiral molecule. With these assumptions, the Hamiltonian simplifies to
\begin{multline}
\label{Eq:29}
  H \\= \hbar\begin{pmatrix} 2n\omega\!+\!\tfrac{1}{2} \omega_\mathrm{M} & \sqrt{n\!+\!1}g^* & \sqrt{n\!+\!1}g^* \\ \sqrt{n\!+\!1}g & (2n\!+\!1)\omega \!-\!\tfrac{1}{2} \omega_\mathrm{M} & 0 \\ \sqrt{n\!+\!1}g & 0 &  (2n\!+\!1)\omega\!-\!\tfrac{1}{2} \omega_\mathrm{M} \\
\end{pmatrix}
\end{multline}
To find its eigenvalues, we introduce the collective field states
\begin{equation}
    \lvert \mathrm{F}_{\pm} \rangle = \frac{1}{\sqrt{2}}\bigl( \lvert n+1 \rangle \lvert n \rangle \pm \lvert n \rangle \lvert n+1 \rangle \bigr)
\end{equation}
Performing a transformation to a new basis $\ket{e}\ket{n}\ket{n},\ket{g}\ket{\mathrm{F}_+},\ket{g}\ket{\mathrm{F}_-}$ using 
\begin{equation}
T = \begin{pmatrix} 1 & 0 & 0 \\ 0& \frac{1}{\sqrt{2}} & -\frac{1}{\sqrt{2}} \\ 0 & \frac{1}{\sqrt{2}} & \frac{1}{\sqrt{2}}  \\
\end{pmatrix}
\end{equation}
the Hamiltonian is given by
\begin{multline}
    H'= T^\mathrm{-1} \cdot H \cdot T \\ =  
    \hbar\begin{pmatrix} 2n\omega\!+\!\tfrac{1}{2} \omega_\mathrm{M} & \sqrt{2(n\!+\!1)}g^* & 0\\ \sqrt{2(n\!+\!1)}g & (2n\!+\!1)\omega \!-\!\tfrac{1}{2} \omega_\mathrm{M} & 0 \\ 0 & 0 &  (2n\!+\!1)\omega\!-\!\tfrac{1}{2} \omega_\mathrm{M} \\
\end{pmatrix}.
\end{multline}
We hence see that the collective field states play the role of super- and sub-radiant states where the roles of molecule and field are reversed in comparison to the usual Dicke scenario \cite{Dicke54}. The superradiant field state $\ket{\mathrm{F}_+}$ exhibits an enhanced coupling with the molecule as evidenced by additional factor $\sqrt{2}$ in front of $g$, whereas the subradiant state $\ket{\mathrm{F}_-}$ completely decouples from the molecule.

The eigenvalues of the two-mode system strongly coupled to the molecule can hence be read off from the single-mode results~(\ref{Eq:17}) and (\ref{Eq:18}), leading to
\begin{eqnarray}
\label{Eq:30}
    E_{1,2} &=& \bigl( 2n + \tfrac{1}{2} \bigr) \hbar \omega \pm \tfrac{1}{2} \hbar \, \Omega\\
\label{Eq:34}    
    E_3&=&\bigl( 2n +  1\bigr) \hbar \omega -\tfrac{1}{2}\hbar\omega_\mathrm{M}
\end{eqnarray}
with Rabi frequency
\begin{equation}
\label{Eq:31}
    \Omega= \sqrt{\Delta^2 + 8(n+1){\lvert g \rvert}^2 }
\end{equation}
and a modified coupling angle $\theta$
\begin{equation}
    \tan(2\theta)= -\frac{8(n+1)\vert g\vert^2}{\Delta}.
\end{equation}

\begin{figure}
    \includegraphics[width=1.\columnwidth]{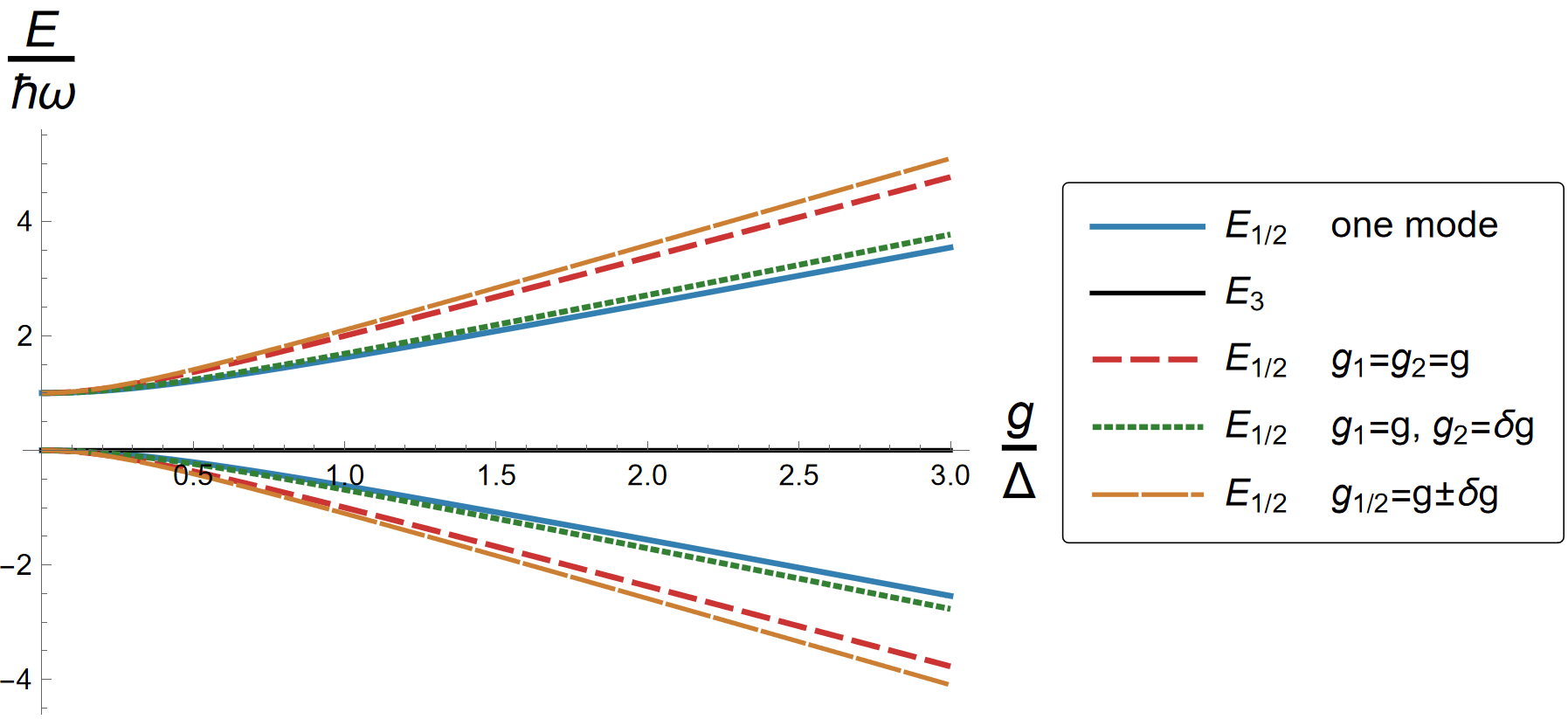}
    \caption{\textit{Eigenenergies for a molecule inside a two-mode cavity.} We display the eigenenergies for zero photon number $n=0$. We show the degenerate case ($g_1=g_2$), the case of an imperfect cavity ($g_1\equiv g$, $g_2\equiv \delta g=0,4g$) and the near-degenerate case ($g_{1,2}=g\pm\delta g$ with $\delta g=0,4 g$). The single-mode case is also shown for comparison.}
    \label{Fig:3}
\end{figure}

The dependence of the eigenenergies of the system on the coupling constant is displayed in Fig.~\ref{Fig:3}. We see that for an achiral molecule, the addition of a second, degenerate mode hence enhances the coupling constant by a factor $\sqrt{2}$ in comparison to the single-mode case with an associated modification of the Rabi frequency. 


\subsection{Nondegenerate case}

The interaction of a chiral molecule with two modes of opposite handedness inside a chiral cavity can be studied analytically for $\omega_1=\omega_2=\omega$, $n_1=n_2=n$, but $g_1\neq g_2$. Diagonalisation of the Hamiltonian~(\ref{Eq:25}) yields the eigenenergies
\begin{eqnarray}
   E_{1,2} &=& \bigl(2n\!+\!\tfrac{1}{2} \bigr) \hbar \omega \pm \tfrac{1}{2} \hbar \, \Omega\\
   E_3&=&(2n\!+\!1)\omega\!-\!\tfrac{1}{2} \omega_\mathrm{M},\\
   \Omega&=&\sqrt{\Delta^2 + 4(n+1)\Bigl({\lvert g_1 \rvert}^2+{\lvert g_2 \rvert}^2\Bigr)}.
\end{eqnarray}
In the following, we discuss the the implications of this solution for two physically relevant scenarios.


\paragraph*{Imperfect single-mode cavity.}

As a first scenario, assume that the cavity mainly supports one of the two possible circularly polarised modes with amplitude $A_0$\cite{Voronin22}, but the other is weakly present due to imperfections of the cavity mirrors with amplitude $\delta A_0\ll A_0$. In this case, we may write
\begin{eqnarray}
\label{Eq:x40}
    g_1&\equiv& g=\frac{\im\omega A_0}{\hbar}(1+\chi)\vec{d}\cdot\vec{v}(\vec{r}),\\
\label{Eq:x41}
g_2&\equiv&\delta g=\frac{\im\omega\delta A_0}{\hbar}(1-\chi)\vec{d}\cdot\vec{v}(\vec{r})\ll g,
\end{eqnarray}
and the Rabi frequency is given by
\begin{equation}
  \Omega=\sqrt{\Delta^2 + 4(n+1)\Bigl({\lvert g \rvert}^2+{\lvert \delta g \rvert}^2\Bigr)}  
\end{equation}
As illustrated in Fig.~\ref{Fig:3}, the presence of the weak mode increases the overall Rabi splitting. However, due to the opposite signs in front of $\chi$ in Eqs.~(\ref{Eq:x40}) and (\ref{Eq:x41}), the second mode reduces the enantio-discriminatory effect.


\paragraph*{Near-degenerate coupling.}

The second case that we want to investigate in more detail is that of a chiral molecule inside a cavity that supports two standing-wave modes of opposite handedness. We may then write $g_1=g+\delta g$ and $g_2=g+\delta g$ where 
\begin{eqnarray}
    g&=&\im\omega A_0\vec{d}\cdot\vec{v}(\vec{r}),\\
    \delta g&=&\im\omega A_0\chi\vec{d}\cdot\vec{v}(\vec{r})\ll g,
\end{eqnarray}
$\omega_1=\omega_2=\omega$, $n_1=n_2=n$.
where the Rabi frequency assumes the form
\begin{eqnarray}
  \Omega&=&\sqrt{\Delta^2 + 4(n+1)\Bigl({\lvert g+\delta g \rvert}^2+{\lvert g-\delta g \rvert}^2\Bigr)}\nonumber\\
  &=&\sqrt{\Delta^2 + 8(n+1)\Bigl({\lvert g \rvert}^2+{\lvert \delta g \rvert}^2\Bigr)}.  
\end{eqnarray}
This result is illustrated in Fig.~\ref{Fig:3}, where the Rabi splitting is increased in comparison with the degenerate two-mode case. However, due to absence of a term linear in $\delta g$, the Rabi splitting is independent of the handedness of the molecule. This reflects the symmtry of the cavity with respect to the two modes of opposite handedness. An enantiodiscriminatory effect could be induced by introducing a second species of optically active molecules which induce a frequency shift between the two modes ($\omega_1\neq\omega_2$) \cite{Bennett20}.
    

\section{Discussion}
\label{Sec:Discussion}

To estimate the order of magnitude of the Rabi frequency and its predicted discriminatory component, we consider these quantities in the case of zero detuning. According to Eq.~(\ref{Eq:18}) and (\ref{Eq:20}, the purely electric, nondiscriminatory vacuum Rabi splitting can then be estimated by ($\chi=0$)
\begin{equation}
\label{Eq:46}
    \Omega=2|g|=\frac{2\omega A_0|d|}{\sqrt{3}}=\sqrt{\frac{2\omega|d|^2}{3\varepsilon_0\hbar V}},
\end{equation}
while the discriminatory component follows upon recalling Eq.~(\ref{Eq:6})
\begin{equation}
\label{Eq:47}
    \Delta\Omega=\sqrt{\frac{2\omega R}{3\varepsilon_0\hbar cV}}.
\end{equation}
\begin{table}
    \centering
     \caption{Rabi frequencies $\Omega$ and chiral shifts $\Delta\Omega$ of exemplary chiral molecules in three different frequency regions.}
    \begin{tabular}{ccccccc}
        \vspace{-0.2cm}\\
         \ & $\nu$/Hz & $d$/D & V/m$^3\ {}^a$ & $\Omega$/Hz & $\Delta\Omega$/Hz &  $P^b$\\ 
         \vspace{-0.25cm}\\ \hline\\
         \vspace{-.6cm}\\
        PO & \ $1.0~10^{11}$ \ & \ 1.72$^c$ \ & $3.4~10^{-9}$ & \ $2.1~10^{3}$ \  & \ $2.1~10^{2}$ \ & \ $4.8~10^{8}$ \ \\
        ME & $3.2~10^{13}$  &  0.17$^d$ & $1.1~10^{-16}$ &  $2.1~10^{7}$  & $2.1~10^{6}$ & $1.5~10^{7}$\\
        FN & $1.8~10^{15}$  &  0.77$^e$ & $6.1~10^{-22}$ &  $2.9~10^{11}$  & $2.9~10^{10}$ & $6.1~10^{4}$ \\
        \vspace{-0.2cm}\\ \hline\\
    \end{tabular}
    \vspace{-0.2cm}\\
    \parbox{\linewidth}{\footnotesize\raggedright PO: propylene oxide,  
        ME: \textit{trans}-methanediol, FN: fenchone\\
        \vspace{0.2cm}
        $^a$ minimal cavity mode volume V = ($\lambda$/2)$^3$ \\
        $^b$ $P=\nu$/$\Delta\Omega$ spectral resolution power to resolve chiral shift\\ 
        $^c$ permanent dipole component $\mu_b$ \cite{Stahl21b}\\
        $^d$ vibrational dipole moment of CH stretch at 3001 cm$^{-1}$ \cite{Franke23}\\
        $^e$  oscillator strength of 3d transition at 7.32 eV \cite{Devlin97} }
    \label{tab:my_label}
\end{table}

Table~\ref{tab:my_label} gives the values for the Rabi splitting and its discriminatory component for three exemplary chiral molecules and the respective cavities. Propylen oxide, also known as methyl oxirane, has been extensively studied over a wide spectral range, including microwave and terahertz spectra \cite{Mesko17, Stahl21a}, mid-infrared vibrational spectra \cite{Vavra24, Barone14}, and optical spectra using circularly polarized light \cite{Contini07, Garcia13}. Propylen oxide has two enantiomeric forms which are the only stable conformers present at room temperature. This, along with a strong permanent dipole moment $\mu_b = 1.72$ D along the $b$-molecular axis, makes it a candidate well suited for studies of molecular chirality. Despite its strong dipole moment, the Rabi frequency is only 2.4 kHz, although we have chosen the smallest possible mode volume V = ($\lambda$/2)$^3$. To calculate the chiral shift $\Delta\Omega$ from Eq.~(\ref{Eq:47}), we have assumed the magnetic moment to be about 100 times smaller than the electrical dipole moment, which is a reasonable assumption for most molecules. The resulting shift of 240 Hz is too small to be spectrocopically resolved even with high resolution millimeter wave spectrometer of resolution power of $P=\Delta\nu/\nu=10^8$.

As an example for the mid-infrared we chose \textit{trans}-methyldiol, CH$_2$(OH)$_2$, which also has an achiral \textit{cis}-conformer, separated by a potential barrier of moderate height. The infrared spectroscopic properties of \textit{trans}-methyldiol have been studied recently (see e.g.\ Refs.~\cite{Jian21, Chen22}) and the infrared band intensities were derived from high-level \textit{ab initio} calculations \cite{Franke23}. For illustration, we chose the CH-stretch mode at 3001 cm$^{-1}$, which possesses a moderate band intensity of 40.62 km/mole, which corresponds to a dipole moment of $d$ = 0.17 D. Due to a much smaller minimal mode volume compared to the millimeter-wave region, the Rabi frequency and the correponding chiral shift $\Delta\Omega$ = 2.1 MHz is more pronouced and can be spectrocopically resolved by high resolution infrared spectrometers.   Moreover, the infrared spectrum of ME is rotationally resolved at typical resolution powers of $P = 10^{7}$ and individual rovibrational lines can be addressed, supporting the assumption of a nearly ideal two-level system.

As an example for the optical frequency range we present fenchone, which has been studied extensively by means of photo-electron circular dichroism (see e.g. Ref.~\cite{Kastner17}). The $3\mathrm{sp}$($\mathrm{d}$) Rydberg states have recently been studied \cite{Singh20}. Oscillator strengths of fenchone have been calculated \cite{Devlin97}, whose value $f$ = 0.0164 we used to derive the dipole moment of 0.77 D for the optical $3\mathrm{d}$ transition at 7.32eV. The very small minimum mode volume of 6.1~10$^{-22}$m$^3$ results in a chiral shift of 29 GHz, which can already be detected using a spectrometer with a low spectral resolution of 6.1~10$^4$. Although this condition can easily be fulfilled by standard spectrometers, the spectral resolution is too low to resolve individual energy states spectrally and to assume a two-level system.    


\section{Conclusion}
\label{Sec:Conclusion}

We have studied the strong coupling of a chiral molecule with one or two circularly polarised standing-wave modes in a chiral cavity. For the single-mode case, we find that the Rabi frequency of the coupled system depends on the relative handedness of molecule and cavity mode and can hence used to discriminate enantiomers. We have further shown that the strong coupling leads to a Casimir--Polder force along the cavity axis whose sign again depends on the handedness of the molecule.

For the more realistic case of two circularly polarised modes being supported by the cavity, we have considered three scenarios. In the case of fully degenerate coupling constants for an achiral molecule, the Rabi splitting is enhanced in comparison to the single-mode case, but becomes entirely nonchiral. The symmetrically displaced coupling constants found for a chiral molecule inside a cavity with equally strong left- and right-handed modes is enhanced in comparison with the degenerate case, but again does not exhibit any enantio-discrimination. In contrast, an asymmetric cavity where a mode of given handedness is supplemented by a weak mode of opposite handedness due to imperfections does retain enantio-discrimination. The effect is slightly reduced due to the impact of the weak mode of opposite handedness.

We found that the Rabi frequency shift in the millimeter wavelength range is too small to be resolved by standard spectroscopy techniques. In contrast, the chiral shift in the optical range is much stronger, but due to the high spectral density, individual molecular energy states are no longer distinguishable and the assumption of a simple two-level system is no longer justified. Our investigations show that the mid-infrared range yields sufficiently large Rabi frequency shifts and that the resolution for distinguishing individual rotational-vibrational levels is sufficiently high.  Therefore, an infrared cavity experiment could be a promising approach to experimentally test the non-destructive method for measuring molecular chirality proposed here.   

The idealised investigations of this work can be generalised by including further couplings to the full many-mode cavity spectrum or by accounting for collective, superradiance effects for molecular ensembles \cite{delPino15,George16} which can enhance both electric and discriminatory molecule--field couplings. In the latter case, the superstrong coupling regime may be reached, necessitating more accurate models that go beyond the rotating-wave approximation. In addition, it would be of interest to consider scenarios where the relation between the handedness and emission direction of circularly polarised light is resolved.


\acknowledgements{We thank Steve M. Barnett and Christian Sch\"{a}fer for discussions. This work has been funded by the German Research Foundation (DFG, Project No.~328961117-CRC 1319 ELCH).}


\bibliography{Bib.bib}

\end{document}